\documentstyle[prl,aps,epsf]{revtex} \def\narrowtext{} \tighten \twocolumn
\input epsf.sty 
\begin{document}
\def\lsim{\buildrel <\over\sim }
 
\title{Destruction of the Fermi Surface in Underdoped
High $T_c$ Superconductors}
\author{
        M. R. Norman,$^1$
        H. Ding,$^{1,2}$
        M. Randeria,$^3$
        J. C. Campuzano,$^{1,2}$
        T. Yokoya,$^4$
        T. Takeuchi,$^{5}$
        T. Takahashi,$^4$
        T. Mochiku,$^6$
        K. Kadowaki,$^7$
        P. Guptasarma,$^1$
        and D. G. Hinks$^1$
       }
\address{
         (1) Materials Science Division, Argonne National Laboratory,
             Argonne, IL 60439 \\
         (2) Department of Physics, University of Illinois at Chicago,
             Chicago, IL 60607\\
         (3) Tata Institute of Fundamental Research, Mumbai 400005, India\\
         (4) Department of Physics, Tohoku University, 980 Sendai, Japan\\
         (5) Department of Crystalline Materials Science, Nagoya 
             University, Nagoya 464-01, Japan\\
         (6) National Research Institute for Metals, Sengen, Tsukuba,
             Ibaraki 305, Japan\\
         (7) Institute of Materials Science, University of Tsukuba, 
             Ibaraki 305, Japan\\
         }

\maketitle
\narrowtext

The Fermi surface, the locus in momentum space of gapless excitations,
is a central concept in the theory of
metals. Even though the optimally doped high temperature 
superconductors exhibit an anomalous normal state, 
angle resolved photoemission spectroscopy (ARPES) has revealed a large Fermi 
surface\cite{CAMPUZANO,OLSON,NK} despite the absence of well-defined 
elementary excitations (quasiparticles) above $T_c$.
However, the even more unusual behavior in the underdoped high temperature  
superconductors, which show a pseudogap above
$T_c$\cite{MARSHALL,NATURE,LOESER}, requires us to carefully re-examine this
concept.  Here, we present the first results on how
the Fermi surface is destroyed as a function of temperature in
underdoped Bi$_2$Sr$_2$CaCu$_2$O$_{8+\delta}$ (Bi2212) using ARPES. 
We find the remarkable effect that different {\bf k} points become 
gapped at different temperatures. 
This leads to a break up of the Fermi surface at a temperature
$T^*$ into disconnected Fermi arcs which shrink with decreasing $T$,
eventually collapsing to the point 
nodes of the $d_{x^2-y^2}$ superconducting ground state below $T_c$.
This novel behavior, where the Fermi surface does not form a continuous contour
in momentum space as in conventional metals, is unprecedented in that it
occurs in the absence of long range order.
Moreover, although the d-wave superconducting gap below $T_c$ smoothly evolves
into the pseudogap above $T_c$, the gaps at different {\bf k} points are
not related to one another above $T_c$ the same way as they are below,
implying an intimate, but non-trivial relation, between the two.

ARPES probes the occupied part of the electron spectrum,
and for quasi-2D systems its intensity $I({\bf k},\omega)$ is
proportional to the Fermi function $f(\omega)$ 
times the one-electron spectral function $A({\bf k},\omega) $\cite{NK}.
In Fig.~1, the solid curves are ARPES 
spectra for an underdoped 85K sample 
at three {\bf k} points on the Fermi surface (determined above
$T^*$) for various temperatures.
To begin with let us look at the superconducting state data at $T = 14$K.
At each {\bf k} point, the sample spectra are pushed back to positive
binding energy ($\omega < 0$) due to the superconducting gap, and 
we also see a resolution limited peak associated with
a well-defined quasiparticle excitation in the superconducting state.
The superconducting gap, as estimated by the position of the sample
leading edge midpoint, is seen to decrease as one moves
from point a near ${\bar M}$ to b to c, closer to the diagonal
$\Gamma-Y$ direction, consistent with a $d_{x^2-y^2}$ order parameter.
Next, consider the changes in Fig.~1 as a function of increasing
$T$. At each {\bf k} point the quasiparticle peak disappears
above $T_c$, but the suppression of spectral
weight -- the pseudogap -- persists well above $T_c$, 
as noted in earlier work\cite{MARSHALL,NATURE,LOESER}.

The striking new feature which is apparent from Fig.~1 is that
the pseudogap at different {\bf k} points closes at different temperatures,
with larger gaps persisting to higher $T$'s.
At point a, near ${\bar M}$, there is a pseudogap at all $T$'s below 
180K, at which the Bi2212 leading edge matches that of Pt.
We take this as the definition of $T^*$ \cite{NATURE} above which the
the largest pseudogap has vanished within the resolution of our 
experiment, and a closed contour of gapless
excitations -- a Fermi surface --  is obtained \cite{DING97}. 
The surprise is that if we move along this Fermi surface to point b
the sample leading edge matches Pt at 120K, which is smaller than $T^*$.
Continuing to point c, about halfway to the diagonal direction, 
we find that the Bi2212 and Pt leading edges match at an even lower 
temperature of 95K.  In addition, we have measured spectra on the
same sample along the Fermi contour
near the $\Gamma Y$ line 
and found no gap at any $T$, even below $T_c$, consistent
with $d_{x^2-y^2}$ anisotropy.

One simple way to quantify the behavior of the gap is to plot the
midpoint of the leading edge of the spectrum (Fig.~1e).
We will say that the pseudogap has closed at a {\bf k} point when 
the midpoint equals zero energy, in accordance with the discussion above.  
From this plot, we find that the pseudogap closes
at point a at a $T$ above 180K, at point b at 120 K, and at point c
just below 95 K.  If we now view these data as a function of decreasing
$T$, the picture of Fig.~2 clearly emerges. The pseudogap suppression
first opens up near $(\pi,0)$ and progressively gaps out larger portions of the
Fermi contour, leading to gapless arcs which shrink with decreasing $T$.
It is worth noting that midpoints with negative binding energy,
particularly for {\bf k} point c, indicate
the formation of a peak in the spectral function at $\omega = 0$ as $T$
increases. 

We see similar results on other underdoped samples.  For example, in 
the upper panel of Fig.~3 we show midpoints for a 77K 
underdoped sample at two {\bf k} points shown in the inset, with behavior
very similar to that of the 85K sample of Fig.~1.
Contrast this behavior to that of the more conventional
$T$-dependence of an overdoped 87K sample shown in the lower panel. 
Gaps with different magnitudes, one at a {\bf k} point near ${\bar M}$
and the other halfway towards 
the $\Gamma Y$ direction, go to zero at the same temperature, very
close to $T_c$,
a behavior we have seen in other overdoped samples as well.  This is
in marked contrast with the new results on underdoped samples.  
Further, to show that the negative midpoints at high $T$'s are not 
unusual, we plot those for an 82K overdoped 
sample at the ${\bar M} Y$ Fermi point as filled symbols in the lower panel.
The midpoint goes to zero at about $T_c$
(indicating the absence of a pseudogap above $T_c$ in this sample)
followed by a slower evolution to negative binding energy 
(indicating the formation of a spectral peak, as discussed above).
There are also important differences between the spectral lineshapes 
of the overdoped and underdoped cases.  The underdoped spectra are 
broader above $T_{c}$, as demonstrated by the flatness of the spectra
seen in Fig.~1 at {\bf k} points a and b, and have smaller quasiparticle
peaks below 
$T_{c}$, implying an increase in the strength of the interactions as 
the doping is reduced.

Before discusing the implications of our results, we
introduce a visualization aid that makes these results very transparent.
This symmetrization method, described in the caption of Fig.~4, effectively
eliminates
the Fermi function $f$ from ARPES data and permits us to focus directly
on the spectral function $A$.
We have extensively checked this method, 
and studied in detail the errors introduced by
incorrect determination of the chemical potential or of ${\bf k}_F$
(which lead to spurious narrow features in the symmetrized spectra),
and the effect of the small ($1^\circ$ radius) {\bf k}-window of the 
experiment (which was found to be small).

In Fig.~4 we show symmetrized spectra for the 85K underdoped sample 
corresponding to the raw data of Fig.~1.
To emphasize that the symmetry is put in by hand, we show
the $\omega > 0$ curve as a dotted line. At {\bf k} point a near 
${\bar M}$ the sharp quasiparticle peak disappears above
$T_c$ but a strong pseuodgap suppression, on the same scale as
the superconducting gap, persists all the way up to 180K ($T^*$). 
Moving to panels b and c in Fig.~4 we again see pseudogap 
depressions on the scale of the superconducting gaps at those points, however 
the pseudogap fills up at lower temperatures: 120K at b and 95K at c.  
In panel c, moreover, a spectral peak at zero energy emerges as 
$T$ is raised. All of the conclusions drawn from the raw
data in Figs.~1 and 3 are immediately obvious from the simple
symmetrization analysis of Fig.~4. 

We now discuss why the $T$ dependence of the Fermi arc
is not simply due to inelastic scattering above $T_c$ broadening the d-wave
node.  From Fig.~4, it is apparent that the gap ``fills in'' for 
{\bf k} points a and b as $T$ is raised, whereas it ``closes'' for 
{\bf k} point c since a peak at zero energy emerges.
This can be seen more clearly in 
Fig.~5, where we show symmetrized spectra for a 75K 
underdoped sample at two {\bf k} points (similar to points a and c of 
Fig.~4) as a function of temperature.  For the first point (I), the spectral 
feature at the gap edge does not move with temperature, whereas for the 
latter point (II), it clearly moves in to zero energy.

We now give a brief discussion of the implications of 
our results. A unique feature of ARPES is that it provides
{\bf k}-resolved information. We believe that the unusual 
$T$-dependence of the pseudogap anisotropy will be a very important 
input in reconciling the different crossovers seen in the
pseudogap regime by different probes. The point here
is that each experiment is measuring a {\bf k}-sum weighted with
a different set of {\bf k}-dependent matrix elements or kinematical
factors (e.g., Fermi velocity).  For instance, quantities which involve
the Fermi velocity, like dc resistivity above $T_c$ and the penetration depth
below $T_c$ (superfluid density),
should be sensitive to the region near the $\Gamma Y$ direction,
and would thus be affected by the behavior we see at {\bf k} point c.
Other types of measurements (e.g. specific heat and tunneling) are
more ``zone-averaged" and will have
significant contributions from {\bf k} points a and b as well,
thus they should see a more pronounced pseudogap effect.
Interestingly, other data we have indicate that the region in the
Brillouin zone where
behavior like {\bf k} point c is seen shrinks as the doping is reduced,
and thus appears to be correlated with the loss of
superfluid density\cite{UEMURA}.
Further, we speculate that the disconnected Fermi arcs 
should have a profound influence on magnetotransport given the lack
of a continuous Fermi contour in momentum space.

We emphasize that the Fermi arcs do {\it not} imply the existence
of hole pockets (i.e., small closed contours) centered about $(\pi/2,\pi/2)$,
as suggested by some theories of lightly doped Mott insulators\cite{LEE}.
In the samples studied here (and more heavily underdoped ones) 
we have carefully searched for hole pockets and for
shadow band dispersion, but found no evidence for either
\cite{DING97}. The gapless arcs that we observe are simply an
intermediate state in the smooth evolution of $d$-wave nodes 
into a full Fermi surface. This smooth evolution was carefully checked 
on an 83K underdoped sample where a detailed sweep was done in {\bf k} space
at $T=90$K, revealing only a small Fermi arc just above $T_c$.
This behavior is fully
consistent with the gap above and below $T_c$ being of the same 
origin as suggested by our earlier experiments
\cite{NATURE,DING97}.

Theoretical calculations in which d-wave pairing correlations cause a
pseudogap above $T_c$ \cite{NAZ} have predicted gapless arcs
which expand as $T$ increases.  Resonating valence bond theories also
lead to gapless arcs above $T_c$ due to spinon pairing \cite{WEN}.
There are other proposals in which the pseudogap has a completely 
different (non-pairing) origin from the superconducting gap.
Given the smoooth evolution we find through $T_c$, they appear difficult
to reconcile with our results.

We thank J. Sadleir and A. Kaminski for their help.
This work was supported by the US Dept of Energy
Basic Energy Sciences,
the US National Science Foundation (NSF),
the NSF Science and Technology Center for
Superconductivity, the CREST of JST, and 
the Ministry of Education, Science and Culture of Japan.
The Synchrotron Radiation Center is supported by the NSF.

\begin{figure}
\epsfxsize=3.4in
\epsfbox{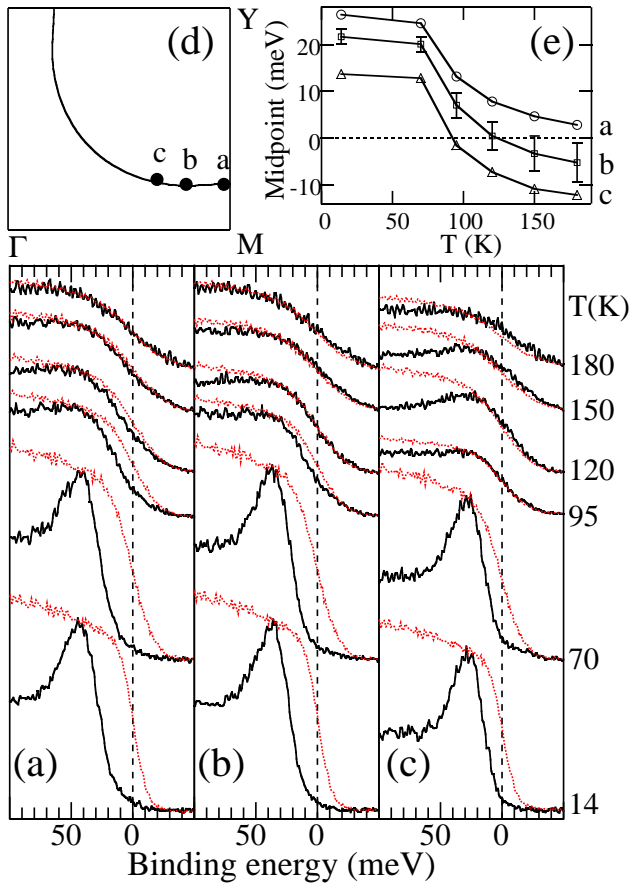}
\vspace{0.5cm}
\caption{
Data obtained on single crystals of Bi2212 grown by the traveling solvent
floating
zone method.  Doping was achieved by adjusting the oxygen partial pressure
during annealing with samples labeled by their onset $T_c$'s.  Measurements were
carried out at the Synchrotron Radiation Center, Wisconsin, using a high
resolution 4-m normal incidence monochromator with 22eV photons and an energy
resolution of 20 meV (FWHM).  The spectra in (a)-(c) are
taken at three ${\bf k}$ points in the
Brillouin zone, shown in (d), for an 85K
underdoped Bi2212 sample at various temperatures (solid curves).
(The $Y$ quadrant was studied to minimize effects due to the
superlattice\protect\cite{GAP}).
Our notation is $\Gamma = (0,0)$,
${\bar M} = (\pi,0)$, and $Y=(\pi,\pi)$, in units of $1/a$, where $a$ is the
Cu-Cu distance, and $\Gamma{\bar M}$ is along the CuO bond direction.
The dotted curves are reference spectra from polycrystalline Pt (in electrical
contact with the sample) used to 
determine the chemical potential (zero binding energy).  Note the closing of the
spectral gap at different $T$ for different ${\bf k}$.  This feature is also
apparent in the plot (e) of the midpoint of the leading edge of the spectra as a
function of $T$.}
\label{fig1}
\end{figure}

\begin{figure}
\epsfxsize=3.0in
\epsfbox{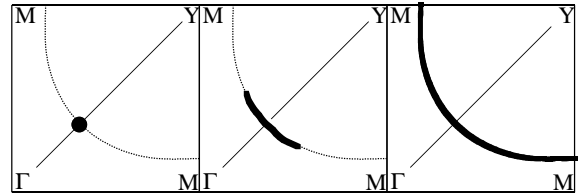}
\vspace{0.5cm}
\caption{
Schematic illustration of the temperature evolution of the 
Fermi surface in underdoped cuprates.
The d-wave node below $T_c$ (left panel)
becomes a gapless arc above $T_c$ (middle panel) which
expands with increasing $T$ to form the
full Fermi surface at $T^*$ (right panel).}
\label{fig2}
\end{figure}

\begin{figure}
\epsfxsize=3.0in
\epsfbox{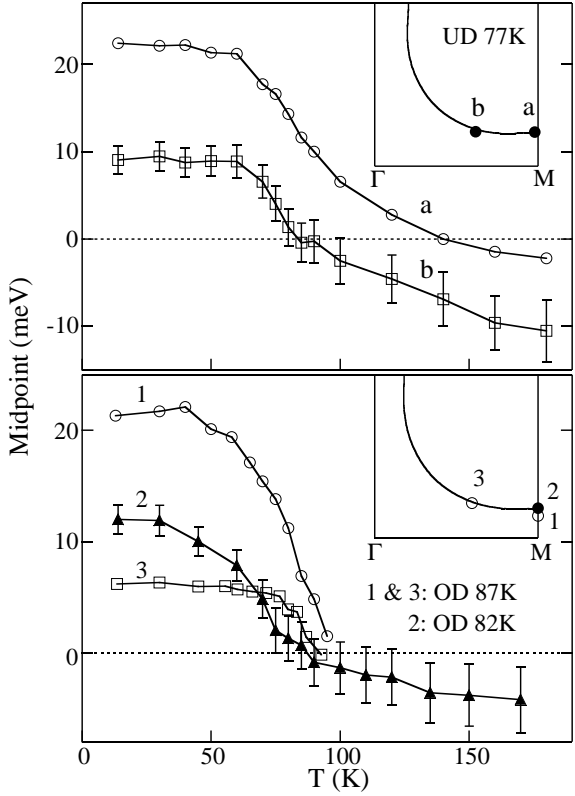}
\vspace{0.5cm}
\caption{
Midpoints of the leading edge of the spectra for a 77K underdoped sample
versus temperature (top
panel), again showing closure of the spectral gap at different $T$ for different
{\bf k}.  This behavior can be contrasted with that of overdoped samples (bottom
panel) where all gaps close near $T_c$.}
\label{fig3}
\end{figure}

\begin{figure}
\epsfxsize=3.4in
\epsfbox{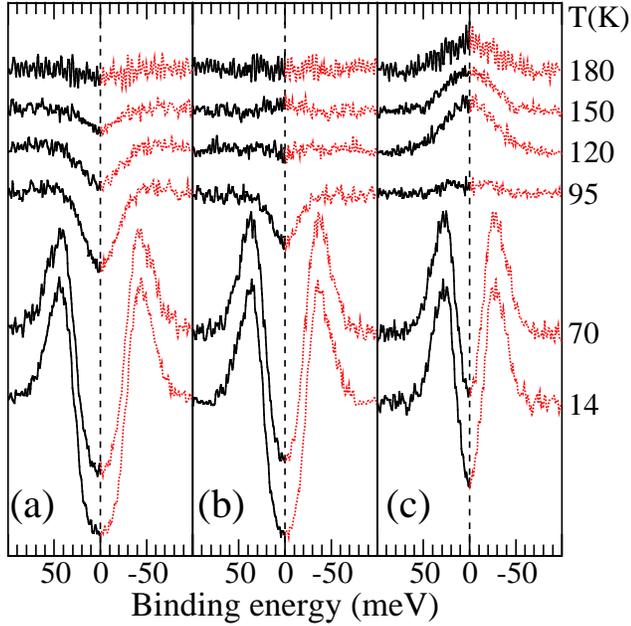}
\vspace{0.5cm}
\caption{
Given ARPES data described by\protect\cite{NK}
$I(\omega)= \sum_{\bf k} I_0 f(\omega)A({\bf k},\omega)$ (with
the sum over a small momentum window about the Fermi momentum ${\bf k}_F$),
we can generate the symmetrized spectrum $I(\omega) + I(-\omega)$.
Making the reasonable assumption of particle-hole (p-h) symmetry 
for a small range of $\omega$ and $\epsilon_{\bf k}$, we have
$A(\epsilon_{\bf k},\omega)=A(-\epsilon_{\bf k},-\omega)$ 
for $|\omega|,|\epsilon|$ less than few tens of meV.
It then follows, using the identity $f(-\omega) = 1-f(\omega)$,
that $I(\omega) + I(-\omega) =  \sum_{\bf k} I_0 A({\bf k},\omega)$
which is true even after convolution with a (symmetric) energy resolution
function.
This symmetrized spectrum coincides with the 
raw data for $\omega \lsim -2.2T_{eff}$, where $4.4T_{eff}$ is
the 10\%-90\% width of the Pt leading edge, which
includes the effects of both temperature and resolution.
Non-trivial information is obtained for the range
$|\omega| \lsim 2.2T_{eff}$, which is then the scale on which p-h
symmetry has to be valid. 
The curves are symmetrized spectra corresponding to the raw spectra of Fig.~1.
The gap closing in the raw spectrum of Fig.~1 corresponds to where
the pseudogap depression disappears in the symmetrized spectrum.
Note the appearance of a spectral peak at higher temperatures in c.}
\label{fig4}
\end{figure}

\begin{figure}
\epsfxsize=3.0in
\epsfbox{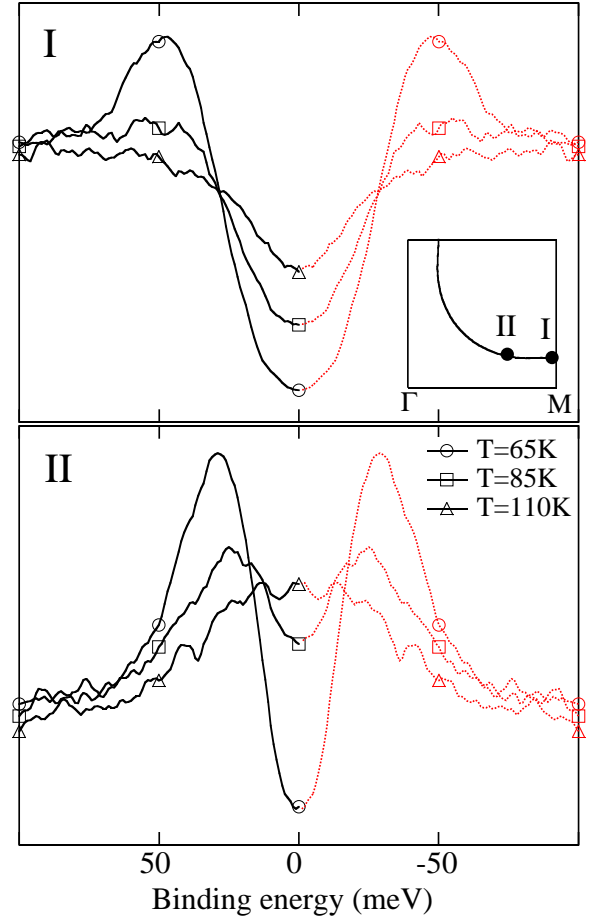}
\vspace{0.5cm}
\caption{
Symmetrized spectra for a 75K underdoped sample for {\bf k} points 
analogous to a and c of Fig.~4 at three different temperatures.
Note that the spectral feature at the gap edge does not move in energy with 
increasing $T$ for 
point I (upper panel), but does move in to zero energy for 
point II (lower panel).}
\label{fig5}
\end{figure}

\end{document}